\documentclass[journal]{IEEEtran}

\usepackage{cite}
\usepackage{amsmath}
\usepackage{array}
\usepackage{booktabs}
\usepackage{graphicx}
\usepackage{float}
\usepackage{subfig}
\usepackage{color}
\usepackage{amsfonts}

\begin{document}
	
\title{Rydberg Atomic Receivers for Net-Zero 6G Wireless Communication and Sensing: Progress, Experiments, and Sustainable Prospects}
	
\author{Yi~Tao,
Zhen~Gao,
Zhiao~Zhu,
De~Mi,~\IEEEmembership{Senior Member, IEEE},
Zhonghuai~Wu,
Zijian~Zhang,~\IEEEmembership{Graduate Student Member,~IEEE},
Fusang~Zhang,
Dezhi~Zheng,~\IEEEmembership{Member,~IEEE},
and Sheng~Chen,~\IEEEmembership{Life Fellow,~IEEE}

\thanks{This work was supported in part by the Natural Science Foundation of China under Grant 62471036; in part by the Shandong Province Natural Science Foundation under Grant ZR2025QA30; and in part by the Beijing Natural Science Foundation under Grants L242011, QY24167, QY25256, QY25257.}
\thanks{
Yi Tao, Zhen Gao (\textit{corresponding author}), Zhiao Zhu, Zhonghuai Wu, and Dezhi Zheng are with the Beijing Institute of Technology, China; De Mi is with Birmingham City University, U.K.; Zijian Zhang is with Tsinghua University, China; Fusang Zhang is with Beihang University, China; Sheng Chen is with the Ocean University of China, China.}
}

\maketitle

\begin{abstract}
Against the backdrop of the global drive to advance the green transformation of the information and communications technology (ICT) industry and leverage technological innovation to facilitate the achievement of Net-Zero carbon goals, research into Rydberg atomic receivers (RAREs) is gaining significant interest.
RAREs leverage the electron transition phenomenon for signal reception, offering significant advantages over conventional radio frequency receivers in terms of miniaturized antenna design, high sensitivity, robust interference resistance, and compact form factors, which positions them as a competitive alternative for meeting zero-carbon communication demands.
This article systematically elaborates on the basic principle, state-of-the-art progress, and novel experiments of RAREs in quantum wireless communication and sensing.
In this first-of-its-kind work, we experimentally verify the RARE-based orthogonal frequency division multiplexing transmission and reveal the potential of deep learning design in optimizing quantum wireless systems.
Finally, we delve into the prospect of integrating RARE with existing cutting-edge application scenarios, while mapping out critical pathways for developing Rydberg-based wireless systems.
\end{abstract}

\vspace{-9mm}
\section{Introduction} 
To mitigate global greenhouse gas emissions, the global community has launched initiatives with the goals of cutting emissions by half by 2030 and achieving Net-Zero carbon emissions by 2050 \cite{ICT}. Meanwhile, the information and communications technology (ICT) industry stands as a key sector for energy conservation and emission reduction. Net-Zero communication is not only crucial to sustainable development but also serves as a vital component of global strategies to address climate change.
However, the enhancement of communication network performance and the industry’s green transition are not without costs. 
The rapid advancement of sixth-generation (6G) networks and artificial intelligence (AI) places extremely high demands on computing power and energy efficiency \cite{zyf}. Moreover, due to the adoption of terahertz (THz) frequency bands, which suffer from severe free-space path loss, 6G networks face significant degradation in signal propagation capability.
To achieve coverage comparable to that of the fifth generation (5G), the number of base stations needs to be increased, leading to exponential growth in energy consumption \cite{ICT}. 
Furthermore, the green transition of the ICT industry also faces technological bottlenecks and cost pressures, all of which pose significant difficulties for advancing Net-Zero communication.
Notably, in the exploration of low-energy solutions for 6G, Rydberg atomic receivers (RAREs) are emerging as a promising direction \cite{magazine}. Based on atomic quantum effects, RAREs leverage the high sensitivity of Rydberg atoms to electric fields, enabling highly sensitive signal reception \cite{MIMO}. Compared with traditional receivers, RAREs exhibit several distinct advantages, as shown below.

\begin{itemize}
\item{
{\bf{Quantum-Limited Sensitivity:}} In ideal conditions, the sensitivity of RAREs can approach the quantum limit. 
Recent advancements in \cite{sensitivity} have demonstrated a noise equivalent power of $9.5 \times 10^{-19} \, \mathrm{W/Hz^{1/2}}$ for the room-temperature THz detector, outperforming conventional systems by four orders of magnitude.
This quantum-limited sensitivity minimizes the energy consumption requirement and enables detection with micro-watt-level laser equipment.
}

\item{
{\bf{Atomic Self-Calibration:}} Conventional metal antennas demand meticulous calibration procedures to achieve uncompromised reception quality.
In contrast, by leveraging atomic standards for self-calibration, RAREs ensure the traceability of measurement results to the international system of units (SI) \cite{s&c}.
This intrinsic stability reduces maintenance overheads, aligning well with the sustainable design requirements of Net-Zero 6G.
}

\item{
{\bf{Wavelength-Independent Miniaturization:}} 
RAREs significantly enhance system integration by replacing metallic antennas and down-conversion modules with atomic vapor cells and photodetectors. Furthermore, Rydberg atoms directly respond to electromagnetic (EM) fields via quantum state manipulation. This decouples the antenna size from the signal wavelength, enabling substantial miniaturization that simplifies hardware architecture and reduces the material-related carbon footprint.
}

\item{
{\bf{Multi-Level to Multi-Band:}} By dynamically manipulating multi-level quantum states of Rydberg atoms, RAREs can achieve full spectrum signal coverage ranging from direct-current to THz frequencies. Replacing multiple radio frequency (RF) chains operating at different frequency bands with this single system enhances flexibility while significantly reducing hardware redundancy and operational energy. Notably, recent work \cite{space-division-multiplexing} has successfully achieved signal reception from 300\,megahertz (MHz) to 25\,gigahertz (GHz) with RARE and facilitated the concurrent processing of dual-band signals.
}

\end{itemize}

\begin{figure*}[!t]
\captionsetup{font={footnotesize}, name = {Fig.}, labelsep = period}
\centering
\includegraphics[width=18cm]{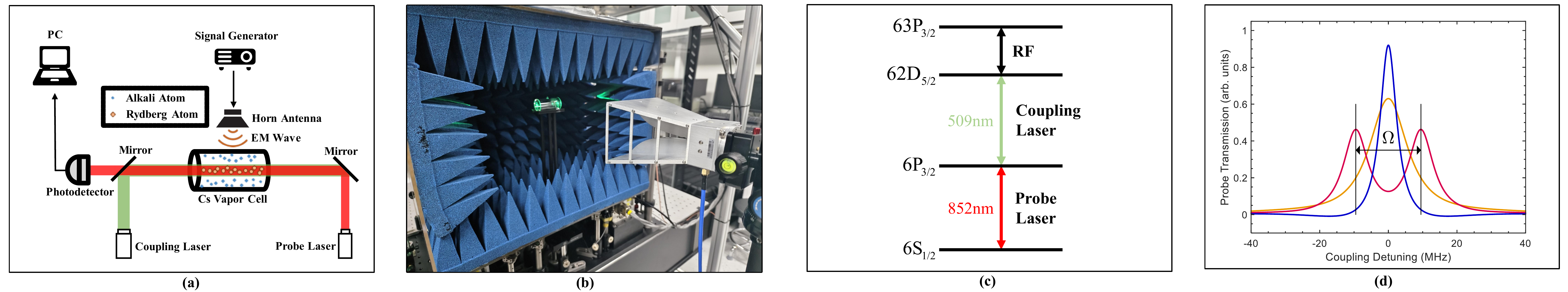}
\caption{Rydberg atom fundamentals. (a) The schematic diagram of an LO-free RARE. (b) The real hardware of the LO-free RARE. (c) The atomic energy level diagram for RF sensing and measurement. (d) The schematic diagram of the Rydberg atom EIT signals and EIT-AT splitting.}
\label{Fig1}
\vspace*{-6mm}
\end{figure*}	 

However, research on RAREs is still exploratory and requires overcoming several key problems.
For instance, in the hardware design, the high cost of lasers presents substantial obstacles to the widespread application \cite{s&c}.
In addition, it is necessary to overcome the challenges of dynamic environmental adaptability, noise suppression, and multi-band processing \cite{MIMO, magazine}.
Fortunately, recent studies have initiated investigations into integrating RAREs with existing, mature wireless solutions.
Recent advancements have highlighted that the implementation of RAREs inherently relies on collaborative design, where the synergistic optimization of hardware architectures and algorithms is indispensable. 

Motivated by this trend, this article provides, for the first time, the experimental verification of 5G/6G-compatible orthogonal frequency-division multiplexing (OFDM) signal reception and demodulation using RAREs.
To align with strategic Net-Zero goals and address the inherent nonlinear response of atomic systems, we employ an AI-driven transmission scheme to achieve cross-layer optimization.
Specifically, we implement RARE-based image transmission experiments using an OFDM waveform and integrate deep joint source channel-coding with quantization (DeepJSCC-Q) \cite{deepjsccq} for enhanced performance, which validates both the feasibility and the superior performance of this proposed system, paving the way for the future development of deep learning design in RAREs.
Finally, we identify key challenges and promising opportunities in wireless communication and sensing, representing the key directions for Net-Zero communication.
\vspace{-2mm}
\section{Preliminary of RAREs} 
An atom is composed of a positively charged nucleus surrounded by negatively charged electrons.
These electrons are distributed in quantized energy level orbits, with each energy level corresponding to a specific energy value. 
Electrons can transition between energy levels by absorbing or emitting photons from EM waves \cite{magazine}. 
When the frequency of a photon resonates with the transition frequency between energy levels, an electronic transition occurs. 
In the case of Rydberg atoms, their outermost electrons are excited to energy states with an exceptionally high principal quantum number $n$. 
This highly excited state results in the electrons occupying orbitals that are significantly distant from the atomic nucleus.
The orbital radius scales proportionally with $n^{2}$, the lifetime of the excited state with $n^{3}$, while the polarizability exhibits a dramatic increase, scaling with $n^{7}$ \cite{MIMO}. 
These distinctive properties make Rydberg atoms exquisitely sensitive to external electric fields, allowing them to detect weak signals that conventional antennas cannot sense, offering a low-power sensing paradigm conducive to Net-Zero goals.

Fig.~\ref{Fig1}\,(a) and Fig.~\ref{Fig1}\,(b) present the schematic diagram and actual photograph of a RARE filled with pure cesium gas, respectively.
The operational principle of this RARE is characterized as ``local oscillator (LO)-free'' \cite{lo-free}. 
Specifically, the RARE establishes a quantum system via a two-photon excitation process.
An energy level can be represented by $ n\ell_j$, in which $\ell$ signifies the orbital angular momentum quantum number and $j$ is the total angular momentum quantum number.
For instance, as shown in Fig.~\ref{Fig1}\,(c), this LO-free RARE comprises the ground state (${\text{6S}_{\text{1/2}}}$), the intermediate state (${\text{6P}_{\text{3/2}}}$), and two Rydberg states (${\text{62D}_{\text{5/2}}}$ and ${\text{63P}_{\text{3/2}}}$). 

First, a weak 852~nm probe laser and a strong 509~nm coupling laser are used to induce the electronic transitions from the ground state to the intermediate state and from the intermediate state to the Rydberg state (${\text{62D}_{\text{5/2}}}$), respectively.
Through this process, ordinary atoms are transformed into Rydberg atoms.
Without the coupling laser, the atomic medium is opaque to the probe laser due to absorption.
However, when the coupling laser is applied and tuned to resonance, it creates a destructive quantum interference, which cancels out the absorption of the probe light, opening a narrow transparency window in the transmission spectrum, known as electromagnetically induced transparency (EIT) \cite{magazine}. This corresponds to the single sharp blue peak shown in Fig.~\ref{Fig1}\,(d), where the transmission is maximized at the resonance frequency.
Since the measurement is based on invariant atomic properties, the theoretical sensitivity is limited only by quantum noise.

Then, a modulated EM wave is transmitted, which necessitates that the frequency is tuned to nearly match the resonance frequency of the transition between the optically excited Rydberg state (${\text{62D}_{\text{5/2}}}$) and a second Rydberg state (${\text{63P}_{\text{3/2}}}$).
Physically, the electric field of the EM wave couples two Rydberg states, which causes the original energy level to split into two distinct energy sublevels, a phenomenon rooted in the alternating current Stark effect \cite{magazine}.
Instead of a single transparency condition at the center frequency, the system now supports two separate transparency conditions corresponding to two new sublevels.
As shown in Fig.~\ref{Fig1}\,(d), as the electric field intensity increases, the original single EIT peak (the single blue peak) first broadens (the orange curve), and finally splits into two symmetric peaks (red curve), which is referred to as Autler-Townes (AT) splitting.
The frequency difference between these two peaks is directly proportional to the signal electric field amplitude $E$, which is expressed as
\begin{align}\label{eq1}
	E = {\hbar\Omega}/{d} ,
\end{align}
where $ d $ is the electric dipole moment of the Rydberg transition, $\hbar = 1.05457 \times 10^{-34} \, \mathrm{J \cdot s}$ is the reduced Planck constant, and $ \Omega $ is the angular Rabi frequency \cite{MIMO}. 
By measuring the spectral splitting of the two peaks, $E$ can be determined, which forms the signal detection principle of RAREs.

\vspace{-3mm}
\section{Review of Existing Rydberg Wireless Communication and Sensing} 

\subsection{Rydberg-Based Wireless Communication}\label{S3.1}

Recent advancements in Rydberg wireless communication include experiments on deep learning-based signal demodulation, video transmission, and theoretical analysis of RARE-enabled multiple-input multiple-output (MIMO) systems.
\begin{itemize}
\item{{\bf{Deep Learning-Based Rydberg Multi-Frequency Signal Demodulation:}}
In 2022, an AI-based scheme was used to combine the RARE for multi-band microwave signal demodulation \cite{deeplearning}. 
In the 4-bin frequency division multiplexing microwave signal with the frequency difference of 2\,kilohertz (kHz), the accuracy reached 99.38\%.
When extended to 20 bins, the RARE was still able to correctly demodulate the data.
Even with the frequency difference increased to 200\,kHz, the accuracy remained at 98.83\%.
Compared to solving the Lindblad master equation, the AI-driven solution demonstrates significant improvements in both accuracy and computational efficiency. This indicates that integrating AI with RAREs can yield high sensitivity and robust interference resistance. Furthermore, the AI-based demodulation scheme effectively reduces computational complexity, which plays a crucial role in minimizing energy consumption.
}

\item{{\bf{Rydberg Video Transmission:}}
In 2022, \cite{TV} demonstrated the first successful transmission of the 480i national television system committee standard format color video signal using the RARE, in which the frame, line, and pixel information were transmitted through a combination of analog waveforms, field and line trigger signals, color burst signals, and a 3.58\,MHz baseband subcarrier (representing the chroma signal).
The signal transmission utilized a carrier frequency of 17.04\,GHz.
The system achieved clear video reception characterized by high-fidelity color reproduction, validating the ability of RAREs to handle complex analog video modulation. Crucially, the ultra-high sensitivity inherent to Rydberg atoms allows for a substantial reduction in the transmit power budget for high-data-rate video services, thereby presenting a prospective energy-efficient solution for future 6G multimedia networks.
}

\item{{\bf{Rydberg MIMO Communication:}}
In 2025, \cite{MIMO} established a comprehensive theoretical framework for RARE-enabled MIMO systems, addressing critical aspects such as signal transmission modeling, channel capacity maximization, and so on.
Specifically, the Rydberg-based signal detection was rigorously characterized as a nonlinear phase retrieval transmission model, and an expectation-maximization (EM) algorithm was developed for iterative signal recovery. 
While this architecture remains in the theoretical model stage at present, the study has still demonstrated the practical feasibility of integrating RAREs into MIMO architectures at the technical level, revealing a potential solution for Net-Zero 6G networks.
}

\end{itemize}

\subsection{Rydberg-Based Wireless Sensing}\label{S3.2}

Current research on RAREs in wireless sensing has notably demonstrated their significant advantages in spectrum sensing and target detection, particularly excelling in enhancing sensing granularity and sensitivity.

\begin{itemize}

\begin{figure}[!t]
	\captionsetup{font={footnotesize}, name = {Fig.}, labelsep = period}
	\centering
	\includegraphics[width=9cm]{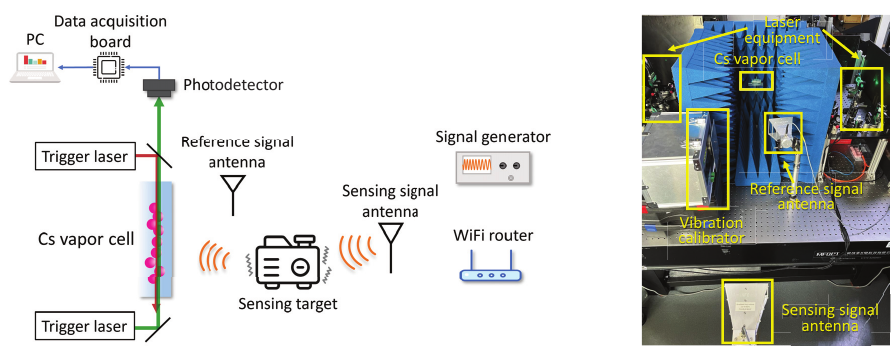}
	\caption{The experimental design of the Rydberg wireless sensing system proposed in \cite{zhangfusang}: the schematic diagram of the Rydberg wireless sensing system and the real hardware of the system.}
	\label{Fig2}
	\vspace*{-5mm}
\end{figure}

\item{{\bf{RARE Enhanced Wireless Sensing:}}
In 2023, \cite{zhangfusang} introduced a quantum wireless sensing system shown in Fig.~\ref{Fig2}, showcasing its ability to advance Wi-Fi sensing granularity from the millimeter (mm) scale to the sub-mm scale, and mm-wave sensing granularity to the micrometer ($\mu$m) scale.
Under a 2.4\,GHz signal, the system reduced the errors in displacement and frequency detection.
Similarly, the system has exhibited exceptional capabilities across the 2.4\,GHz, 5\,GHz, and 28\,GHz frequency bands, consistently delivering improved sensing granularity.
Experiments conducted within a 3-meter distance range between the transmitter and the RARE, or in environments with obstacles, showed that the performance was not significantly affected.
Additionally, the system has withstood ambient RF interference from Wi-Fi and Bluetooth, demonstrating its high interference resilience.
Then, a 2.4\,GHz signal was used to recover music from speaker vibrations, and a 28\,GHz signal to detect $\mu$m-level vibrations in a water bottle caused by footsteps. 
The system accurately reconstructed the speaker's sound with a recovery similarity of 88.8\%, and a 100\% accuracy rate in footsteps detection, outperforming conventional sensing systems and offering an energy-efficient solution for precise sensing in Net-Zero networks.}

\item{{\bf{Rydberg Remote Sensing:}}
In 2024, \cite{satellite} utilized a RARE combined with satellite signals of opportunity to accomplish remote sensing of soil moisture.
The signal waveform is primarily quadrature phase shift keying (QPSK) modulated, covering the 2.32--2.345\,GHz frequency band of the XM satellite.
Weak modulated signals below the atomic readout noise can be extracted via correlation with a known reference waveform, successfully enabling continuous detection of the XM satellite signal spectral envelope. 
Laboratory experiments using simulated XM signals for soil moisture inversion demonstrated an error of less than 0.5\%. 
Outdoor soil moisture sensing using XM satellite signals showed that the Rydberg atomic soil moisture retrievals closely matched the conventional schemes, validating the detection capability of RAREs within ultra-weak signal regimes.}

\end{itemize}

As mentioned above, RAREs are making advancements across multiple domains. These energy-efficient characteristics position RAREs as a key enabler for future Net-Zero 6G networks.
However, the existing Rydberg atom-based system still suffers from several drawbacks.
Specifically, these drawbacks include reliance on customized analog signals, incompatibility with commonly used 5G/6G waveforms, low spectral efficiency, and deviations of theoretical models from complex real-world scenarios, among others. Overcoming these technical bottlenecks is crucial for fully unlocking the potential of RAREs and reducing energy consumption to achieve Net-Zero carbon emissions.

\vspace{-3mm}
\section{AI-Driven Rydberg OFDM Wireless Communication Experiment}\label{S4}
This section presents the experimental verification of the successful reception and demodulation of OFDM waveforms using a RARE and an AI-driven image transmission scheme for improved performance.
Specifically, we introduce the DeepJSCC-Q, which can address the challenge of high-capacity data transmission under bandwidth-constrained scenarios.
Notably, this AI-driven scheme mitigates the inherent nonlinear distortions of the Rydberg channel.
By replacing the computational solution of the Lindblad master equation with efficient neural network inference, it significantly reduces digital processing power consumption.
Combined with the low-power characteristics of the RARE, the proposed scheme provides a low-carbon pathway for advancing the ICT industry toward the goal of Net-Zero 6G.

\vspace{-3mm}
\subsection{System Design}\label{S4.1}
Fundamentally, RAREs are categorized into two distinct architectures, termed LO-free and LO-dressed \cite{lo-free}\footnote{A comprehensive implementation of these two designs is beyond the scope of our discussion, and interested readers may refer to \cite{lo-free} for more details.}. 
In this article, the LO-free RARE is utilized, as shown in Figs. 1(a) and 1(b), which are intended to verify that signal transmission can be achieved even under conditions of lower sensitivity. Meanwhile, LO-dressed architecture can also be utilized.
To overcome the physical limitation of amplitude-only detection, we draw inspiration from the direct-current biased optical OFDM \cite{DCO-OFDM} in optical communications.
{By constructing frequency-domain signals with Hermitian symmetry, we can generate real-valued time-domain signals via the inverse fast Fourier transform, which eliminates the need for in-phase/quadrature (I/Q) orthogonal modulation.}

\begin{figure*}[!t]
	\captionsetup{font={footnotesize}, name = {Fig.}, labelsep = period}
	\centering
	\includegraphics[width=18cm]{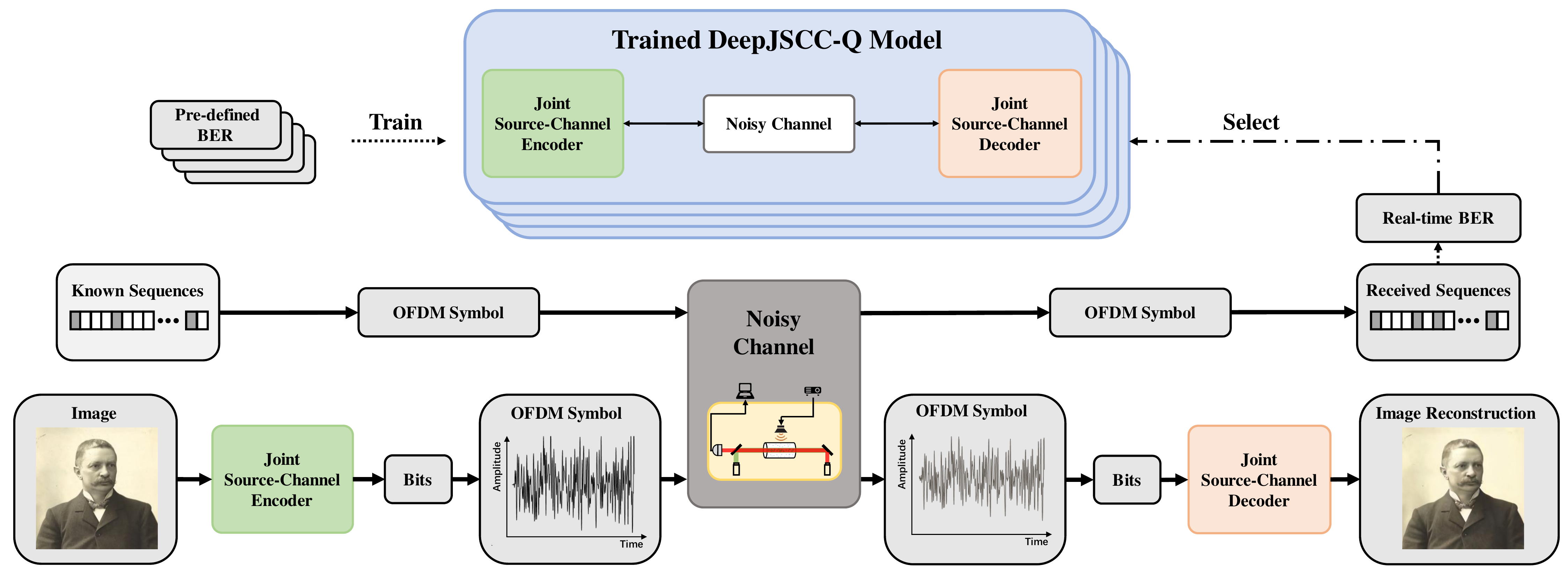}
	\caption{The experimental workflow of the proposed AI-driven Rydberg OFDM wireless communication system. The transmitted image is of Johannes Robert Rydberg (1854-1919), a Swedish physicist and mathematician, and one of the founders of spectroscopy.}
\label{Fig3}
\vspace*{-6mm}
\end{figure*}
We employ the aforementioned OFDM waveform architecture, utilizing quadrature amplitude modulation (QAM) for data mapping.
The carrier frequency is set to 2.911\,GHz, with the energy level selection for the RARE directly referencing the characteristics depicted in Fig. 1(c).
The system operates with an oversampling factor of 4, meaning the effective sampling rate is explicitly four times the baseband signal bandwidth.
Furthermore, the length of the cyclic prefix is set to 1/8 of the total number of subcarriers. 
Pilot design is verified using the comb and block pilots. 
The comb pilots are equally spaced in the frequency domain at intervals of 4 subcarriers. In contrast, the block pilots adopt a time-frequency distribution strategy of inserting one full-symbol pilot every 4 OFDM symbols.
For analysis of system performance, we select 256, 512, and 1024 as configured subcarrier numbers.
The sampling rates are configured at 48\,kHz and 384\,kHz.
Additionally, to mitigate the inherently high peak-to-average power ratio of OFDM waveforms, a hard-clipping scheme with a 5 dB threshold is employed to strictly constrain peak amplitude excursions.
The received signals are processed for channel estimation via the least squares method using pilot symbols, followed by zero-forcing equalization and demodulation. 
\vspace{-3mm}
\subsection{AI Enhanced Rydberg Transmission}\label{S4.2}
Shannon's channel capacity formula reveals the relationship between the maximum transmission rate and the signal power, noise power, and bandwidth, which is represented as
\begin{equation}\label{eq2}
  C = B  \log_2 \left(1 + {S}/{(N + I)}\right), 
\end{equation}
where $C$ is the channel capacity, $B$ is the channel bandwidth, $S$ is the signal power, $N$ is the thermal noise power, and $I$ is the external interference power.
For Rydberg quantum technology, the quantum noise is significantly lower than the thermal noise.
However, in practical applications, environmental interference and the nonlinear response characteristics of RAREs remain core challenges. Fortunately, AI technologies offer a new solution for optimizing system performance, combating noise and interference, and achieving carbon neutrality.

Specifically, we adopt the DeepJSCC-Q proposed in \cite{deepjsccq}.
This architecture integrates image compression, adaptive quantization, channel-aware modulation, and end-to-end reconstruction within a unified framework, significantly enhancing image transmission quality across diverse channel conditions.
Notably, to address the complex nonlinear quantum noise inherent in Rydberg channels, such as laser phase noise and Doppler broadening, the proposed scheme leverages data-driven end-to-end optimization to uniformly quantify these physical impairments as the bit error rate (BER), thereby eliminating the necessity for precise modeling of complex physical processes.
Furthermore, the incorporation of a differentiable quantization layer ensures seamless compatibility with OFDM waveforms, enabling robust adaptability to real-world atomic communication scenarios.

The operational workflow of the proposed AI-driven Rydberg OFDM wireless communication system is illustrated in Fig.~\ref{Fig3}.
During the training process, the DeepJSCC-Q model leverages predefined BERs as parameters, establishing a BER-optimized deep learning model mapping table and enabling optimal image transmission across dynamic channel conditions.
For testing, known bit sequences are transmitted using the designed OFDM waveform and received by the RARE.
The BER is derived from the received sequences and known sequences. 
Leveraging this real-time BER and the established mapping table, the DeepJSCC-Q model that best matches the current channel conditions is selected. 
Subsequently, the system proceeds to the formal image transmission. 
The image undergoes joint source-channel encoding via the encoder, and then it is transformed into OFDM symbols for transmission. 
After being received by the RARE, signals are processed for equalization and demodulation.
Then, joint source-channel decoding is performed to achieve the image reconstruction.

\begin{figure}[!t]
\captionsetup{font={footnotesize}, name = {Fig.}, labelsep = period}
\centering
\includegraphics[width=7.7cm]{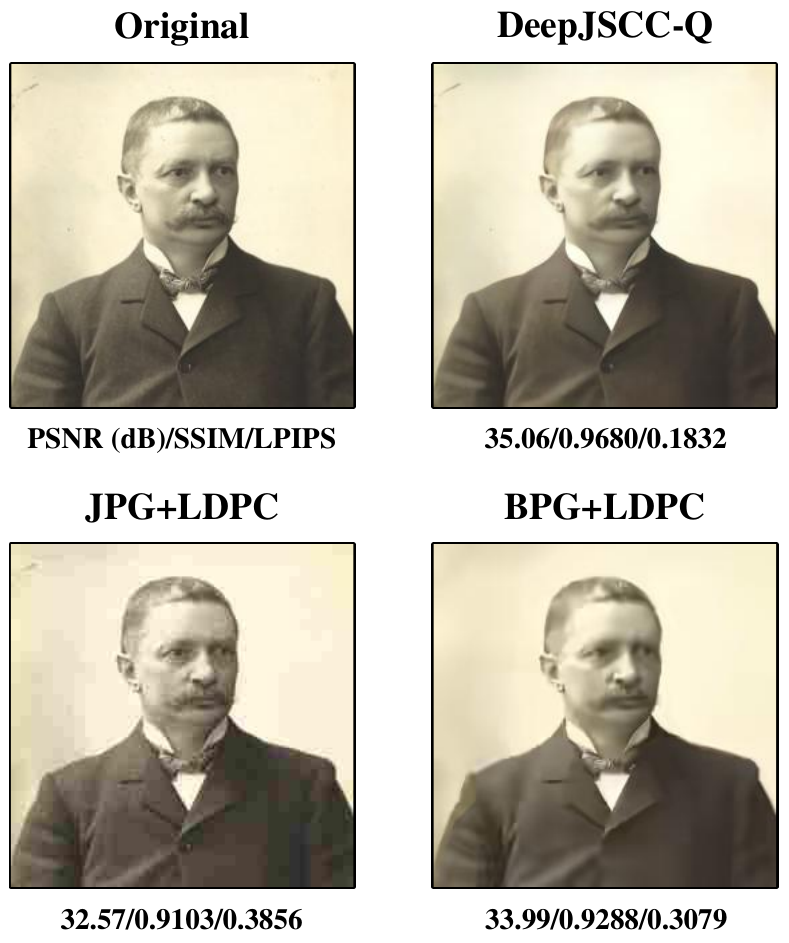}
\caption{Rydberg wireless communication experimental results of the image transmission performance using PSNR, SSIM, and LPIPS for evaluation.}
\label{Fig4}
\vspace*{-5mm}
\end{figure}

\vspace{-2mm}
\subsection{Experimental Results}\label{S4.3}
We observe distinct differences in BER performance across different sampling rates. At 48 kHz, the BER remains at approximately 1\%; however, increasing the sampling rate to 384 kHz degrades the BER to around 7\%.
This is because Rydberg atoms are only sensitive to specific transition frequencies, while OFDM subcarriers are distributed across a wide frequency band. Therefore, most subcarriers are in a non-resonant state due to detuning. 
Moreover, the wide bandwidth causes more noise accumulation, which further degrades the demodulation accuracy.
Additionally, precision limitations of equipment, i.e., potential laser frequency instability, can ultimately induce carrier frequency offsets.
For the number of subcarriers, variations have a relatively minor impact on BER, which is because the average signal powers of different numbers of subcarriers after clipping processing tend to be close. 
Experimental results demonstrate that the BER performance using comb pilots outperforms that of block pilots, indicating the Rydberg channel exhibits significant time-varying characteristics, reflecting the high sensitivity of the RARE to EM fields.

In the experimental comparison shown in Fig.~\ref{Fig4}, we evaluate the performance between DeepJSCC-Q and the two conventional separate source-channel coding schemes without deep learning, based on JPG+low-density parity-check (LDPC) codes and BPG+LDPC codes. The number of subcarriers is 1024, comb pilots are used, and the sampling rate is 384\,kHz.
Across three image quality metrics, i.e., peak signal-to-noise ratio (PSNR), structural similarity index measure (SSIM), and learned perceptual image patch similarity (LPIPS), the experimental results indicate that DeepJSCC-Q demonstrates superior image reconstruction quality. 
By end-to-end processing, this AI-driven image transmission design effectively adapts to the Rydberg channel's characteristics while achieving reduced resource consumption.

\begin{figure*}[!t]
	\captionsetup{font={footnotesize}, name = {Fig.}, labelsep = period}
	\centering
	\includegraphics[width=14.4cm]{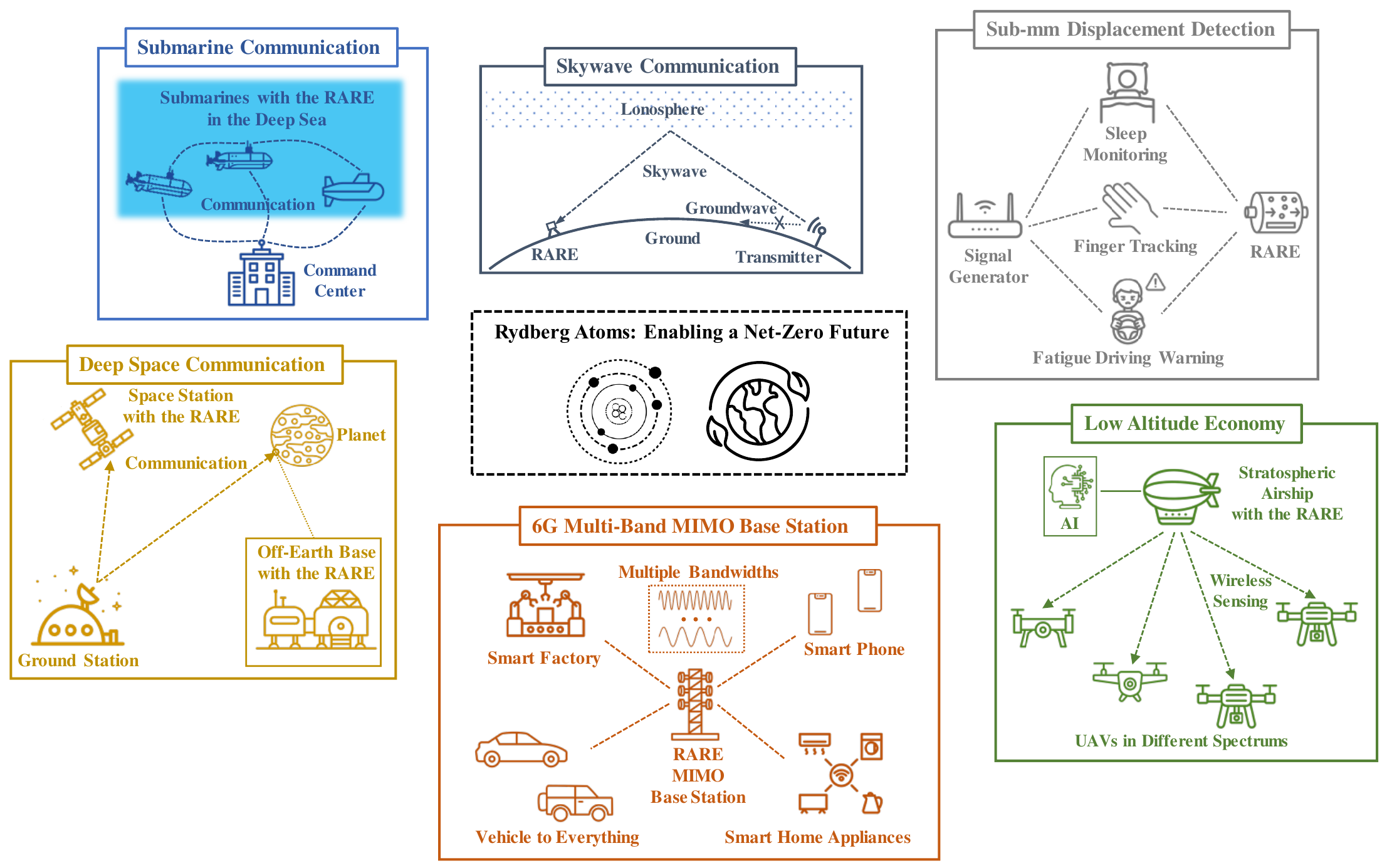}
	\caption{Rydberg technology for Net-Zero enabled future wireless communication and sensing.}
	\label{Fig5}
	\vspace*{-6mm}
\end{figure*}	 

\vspace{-2mm}
\subsection{Discussion and Insights}\label{S4.4}
While the feasibility of AI-based RARE transmission has been verified, significant potential for enhancement remains.
Regarding parameter optimization, adaptive matching between quantization and modulation orders could be established, alongside model architecture refinements. Signal processing, specifically channel estimation and equalization, can be improved by employing advanced neural networks to compensate for complex channel dynamics.
In terms of implementation, integrating LO-dressed modes and other advanced RARE configurations warrants further exploration.
To align with Net-Zero goals, green AI algorithms such as attention-based sparse computing can be adopted to reduce computational and energy costs. 
Furthermore, leveraging physical mechanisms like multi-level electron transitions offers another promising avenue for performance optimization.

\vspace{-4mm}
\section{Prospects and Challenges for Future Rydberg Wireless Communication and Sensing}\label{S5}

In this section, we discuss the prospects and challenges of Rydberg atom-based wireless communication and sensing technologies, focusing primarily on multi-scenario adaptability, practical implementation, and cross-layer integration, while integrating Net-Zero goals to propose technical optimization directions, as depicted in Fig. 5.

\vspace{-3mm}
\subsection{Rydberg Wireless Communication in Over-the-Horizon and Ultra-Long-Range Scenarios}\label{S5.1}

RAREs present a transformative solution for over-the-horizon and ultra-long-range scenarios, including skywave, submarine, and deep space communications.
Conventionally, such systems rely on massive antenna arrays to facilitate low-frequency and long-distance transmission.
However, these large-scale infrastructures suffer from inherent drawbacks, including bulky dimensions, poor interference resistance, and high material consumption during manufacturing.
In contrast, RAREs enable substantial antenna miniaturization.
Quantitatively, unlike conventional antennas constrained by the Chu-Harrington limit to sizes comparable to the wavelength $\lambda$, RAREs are decoupled from $\lambda$.
For instance, at 30\,MHz ($\lambda \approx 10$\,m), a centimeter-scale vapor cell achieves a physical volume reduction compared to a standard half-wave dipole, significantly lowering material carbon footprints.
Moreover, the integration of Rydberg vapor cells with photonic waveguides and micro-electro-mechanical systems further minimizes sensor footprint and operational energy usage, directly aligning with Net-Zero targets.
As demonstrated in \cite{shortwave}, the RARE has achieved a remarkable three-order-of-magnitude improvement in sensitivity and successfully received broadcast signals from 880 kilometers away, verifying the potential of RAREs in over-the-horizon communication.
Furthermore, addressing the severe signal attenuation in ultra-long-range transmission, the intrinsic ultra-high sensitivity of RAREs notably enables the detection of faint signals across vast distances, demonstrating exceptional prospects for future applications.

\vspace{-3mm}
\subsection{{Rydberg Enabling Compact Multi-Band Integrated MIMO Array for 5G/6G}}\label{S5.2}

The 5G spectrum covers the range from 450\,MHz to 6\,GHz and from 24.25\,GHz to 52.6\,GHz, providing services for a variety of application scenarios. 
Looking ahead, 6G is expected to expand into higher frequency bands ranging from 90\,GHz to over 300\,GHz.
Based on the evolution trend of the 5G multi-band antenna architecture, the requirements of 6G ultra-large-scale antenna arrays aim to achieve a leap in multi-band collaborative capabilities. 
Therefore, Rydberg-based multi-band MIMO base stations are expected to revolutionize base station design.

As mentioned above, a single RARE is capable of simultaneously processing multiple frequency bands.
Integrating the multi-band response characteristics with OFDM waveforms enables synergistic optimization of cross-band diversity gain and spectral efficiency, while reducing base station hardware redundancy and operational energy consumption.
Furthermore, building upon the MIMO atomic receiver \cite{MIMO}, RAREs can achieve MIMO-OFDM communications in real-time. 
Although currently Rydberg atoms can only perform spectrum scanning to capture multiple narrowband signals \cite{s&c}, the potential for RAREs to replace multiple conventional receive antennas remains promising.
Additionally, RAREs are expected to simultaneously integrate communication and sensing capabilities across multiple frequency bands.
Therefore, a novel waveform with joint space-time-frequency modulation is expected, providing technical support for multi-band collaboration and low-carbon operation.
In summary, despite the lack of practical Rydberg MIMO architectures, existing verifications of complex waveforms, e.g., OFDM, provide a solid foundation for future research.
Future research efforts for the Rydberg multi-band MIMO base station design must focus on the multi-band solutions, cost reduction, as well as the optimization of cross-band channel modeling and waveform design, to achieve the synergistic development of technical performance and Net-Zero goals.

\vspace{-3mm}
\subsection{Rydberg Wireless Communication and Sensing for Low Altitude Economy}\label{S5.3}

The full-spectrum coverage and miniaturized nature of RAREs offer significant value for high-sensitivity EM environment monitoring in low-altitude economies.
The miniaturized nature of RAREs allows for flexible deployment on high-altitude platforms such as satellites \cite{satellite} and stratospheric airships \cite{s&c}.
The low power consumption of RARE significantly reduces the payload's energy burden, extending operational endurance and aligning with the sustainability requirements of green aviation.
The multi-band coverage capability of RAREs enables the simultaneous capture of various communication and radar signals of different unmanned aerial vehicles (UAVs). 
Additionally, leveraging an AI-driven strategy for signal processing, RAREs are capable of effectively suppressing signal interference, achieving high sensitivity perception to ensure the security of low-altitude airspace.

\vspace{-3mm}
\subsection{Rydberg Wireless Sensing for Sub-mm Detection}\label{S5.4}

Rydberg wireless sensing can enhance the sensing granularity of conventional RF sensing due to its extremely low noise level \cite{zhangfusang}, which is expected to not only achieve sub-mm level displacement detection and bring about revolutionary progress in the field of wireless sensing but also, by virtue of its ultra-low power consumption, provide a pathway for green sensing.
For instance, this technology can accurately capture the mm-level displacement of the chest caused by breathing and the tiny tremors produced by heartbeats, thus enabling continuous non-contact monitoring of breathing frequency and heart rate, effectively solving the problem of conventional contact sensors affecting sleep comfort. 
Moreover, the superior sensing capabilities of RARE promise a significant leap in gesture tracking technology, allowing for the precise tracking of individual finger movements, a marked improvement over the tracking of general hand motions \cite{zhangfusang}. This will greatly improve the accuracy and intuitiveness of human-computer interaction.
Moreover, by monitoring the subtle movements of a driver's head, it can provide an early warning of fatigue driving, enhancing driving safety. 
Nevertheless, higher sensing granularity also means that quantum wireless sensing is more susceptible to interference from non-target movements, which is a challenge that needs to be addressed in future research.

\section{Conclusions}\label{S6}

In this article, in combination with the future Net-Zero communications, we have elaborated on the wireless communication and sensing potential of RAREs, from system design to current developments, experiments, and prospects.
As a cutting-edge platform of quantum precision measurement, RAREs have unique quantum state response mechanisms, which have driven innovations in diversified applications.
This article, for the first time, has presented an efficient end-to-end image transmission experiment under a Rydberg-based quantum receiver and confirmed the potential of integrating the commonly used OFDM waveform and the emerging AI-driven transmission strategy into RAREs.
The image reconstruction performance of the proposed system is significantly better than that of conventional schemes, revealing the critical importance of multi-dimensional joint optimization.
Looking ahead, Rydberg atomic technology is expected to address real-world challenges across multiple scenarios, being applied in over-the-horizon and ultra-long-range communication systems,  ultra-large-scale MIMO base stations in 5G/6G, low-altitude economic applications, and sub-mm displacement detection, while enabling the realization of Net-Zero carbon communications, multi-band signal processing, integrated sensing and communication functionalities, precise sensing, and so on.
	
\ifCLASSOPTIONcaptionsoff
\newpage
\fi

\end{document}